      [1999/12/01 v1.4c Il Nuovo Cimento]
\documentclass{cimento}

\usepackage{graphicx}  % got figures? uncomment this
\title{Are Durations of Weak Gamma-Ray Bursts Reliable?}
\author{Maarten Schmidt}
\instlist{\inst California Institute of Technology}
\PACSes{\PACSit{95.85.Pw}{gamma-ray}
\PACSit{98.70.Rz}{gamma-ray bursts}}

\begin{document}

\maketitle

\begin{abstract}

Simulations in the GUSBAD Catalog of gamma-ray bursts suggest that the
apparent duration of a burst decreases as its amplitude is decreased. 
We see no evidence for this effect in the BATSE catalog. We show that
for a burst at the detection limit, the typical signal-to-noise ratio at
the edges of the $T_{90}$ duration is around 1.5, suggesting that
$T_{90}$ must be quite uncertain. The situation for $T_{50}$ is less
unfavorable. Simulations using the exact procedure to derive the 
durations listed in the BATSE catalog would be useful in quantifying
the effect.
\end{abstract}

\section{Introduction}

Astronomers are generally familiar with the fact that the observation
of spectra of astronomical objects requires many more photons than a
mere detection. For that reason, spectra are usually available for
the brighter objects in a catalog only. A similar consideration of
durations of gamma-ray bursts (GRB) would suggest that these are only
given for the brighter objects in a catalog: it takes more
photons to define the time profile needed to derive the duration than
to just detect the burst. Actually, the BATSE catalog lists durations 
$T_{50}$ and $T_{90}$ for GRBs regardless of peak flux. Bias in the
durations of weak GRBs has been a concern~\cite{ref:n96,ref:h00}.
In this paper we discuss simulations that illustrate the situation 
in deriving durations for weak bursts.

\section{Simulations Based on the GUSBAD Catalog}

Our own experience is based on the GUSBAD catalog~\cite{ref:s03,ref:s04},
which is based on BATSE DISCLA data at a time resolution of 1024 ms.
The catalog covers the full $CGRO$ mission from April 19, 1991 till
May 26, 2000. The detection algorithm required an excess of at least
$5 \sigma$ in at least two detectors in the energy range $50-300$ keV. 
The search was limited to times when incomplete or contaminated data
were not interfering in the detection process. As a consequence, for 
each of the 2204 GRBs in the catalog all derived properties are given.
There are 589 GRBs in the GUSBAD catalog that are not listed in the 
BATSE catalogs.

\begin{figure}
\includegraphics[totalheight=12cm,angle=-90]{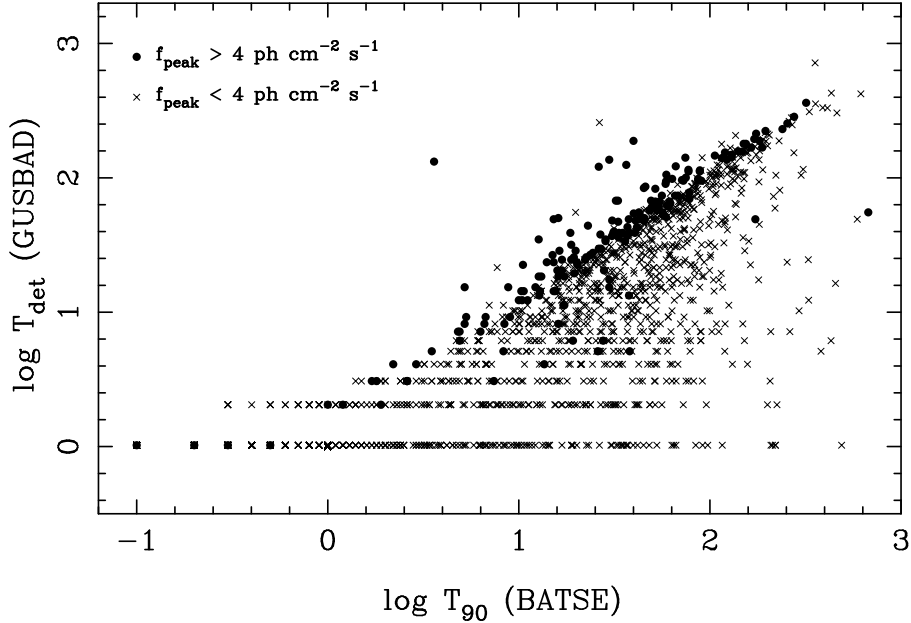}  
\caption{Durations $T_{\rm det}$ derived from the GUSBAD catalog
 plotted against the BATSE duration $T_{90}$.}
\end{figure}

Among the properties given in the GUSBAD catalog are the time bin of
first detection (trigger) and the last time bin when the flux exceeded
the detection flux limit. We do not give durations in the catalog but
one can use these two time bins to define a {\sl detection duration}
$T_{\rm det}$ which is the total time span over which the GRB was detectable.
In the derivation of the Euclidean value of $V/V_{\rm max}$ a simulation 
was carried out in which the (unknown) distance of the GRB was increased
until it did not trigger the software trigger anymore. We noticed in this 
process that as the amplitude of the burst declined, $T_{\rm det}$
became smaller and smaller and would end up at one time bin (1024 ms) 
at the detection limit. Clearly $T_{\rm det}$ for weak GRBs is not a 
physically meaningful duration, which is the reason why we did not 
list it in the GUSBAD catalog. 

One may ask whether the detection time $T_{\rm det}$
is a valid measure of duration for strong GRBs. Figure 1 compares
GUSBAD $T_{\rm det}$ and BATSE $T_{90}$ durations for objects common to
the two catalogs. For strong GRBs there appears to be a good correlation
between the two, with a small number of outliers. For weak GRBs the 
correlation between $T_{\rm det}$ and $T_{90}$ breaks down. For these 
sources, the $T_{90}$ durations range from a fraction of a second to 
several hundred seconds. 

\begin{figure}
\includegraphics[totalheight=12cm,angle=-90]{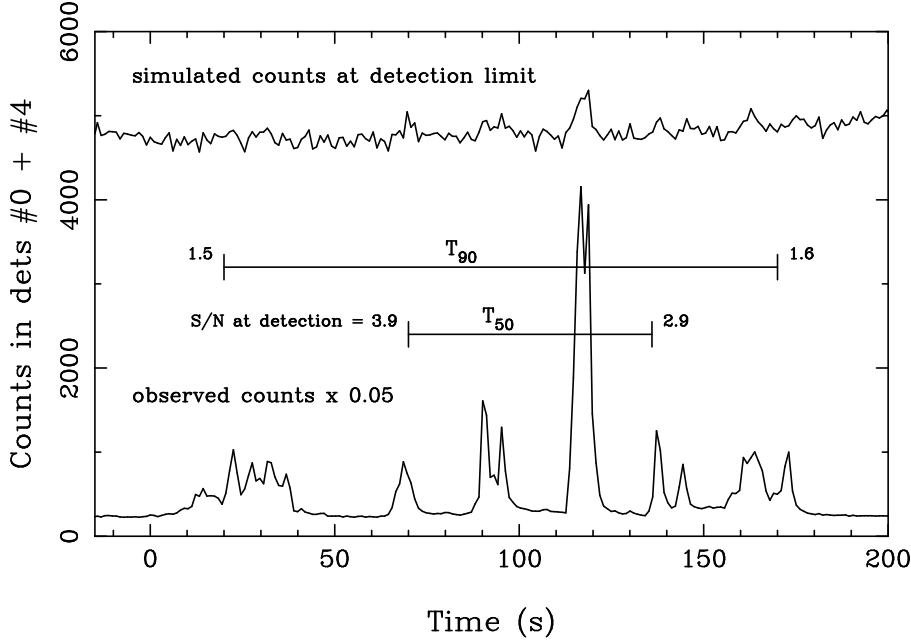}  
\caption{Results of a simulation showing how the strong burst GRB 940217 
would be seen if its amplitude is decreased to be at the BATSE limit
of detection. The observed time profile of the burst is also shown,
together with the $T_{90}$ and $T_{50}$ windows according to the BATSE
catalog.}
\end{figure}

We illustrate the derivation of duration for weak bursts by a simulation
on the strong BATSE GRB 940217 = GUSBAD 940217.960. We show in Figure 2
the sum of the counts in the two brightest illuminated detectors,
reduced by a factor of 20. At the top of the figure, we plot the simulated
counts at the detection limit (corresponding to the BATSE trigger of an 
excess of $5.5 \sigma$ over background in the second detector). These counts 
are the sum of the reduced burst signal and a stretch of background 
preceding the burst. Also shown are the time windows corresponding to 
the two durations given in the BATSE catalog as well as the S/N ratios
at either end of the windows. These S/N ratios are actually averaged
over 24 GRBs common to the BATSE and GUSBAD catalogs with 
$C_{\rm max} > 5000$ and $T_{\rm det} > 100$ s. The average S/N ratio
of the peak is 8.3. The edges of the $T_{90}$ window have S/N ratios
around 1.5, suggesting that the reliability of $T_{90}$ is problematic
at best. The edges of the $T_{50}$ window in the simulation have an 
average S/N ratio around $3-4$ at the detection limit. 

\section{Discussion}

\begin{figure}
\includegraphics[totalheight=12cm,angle=-90]{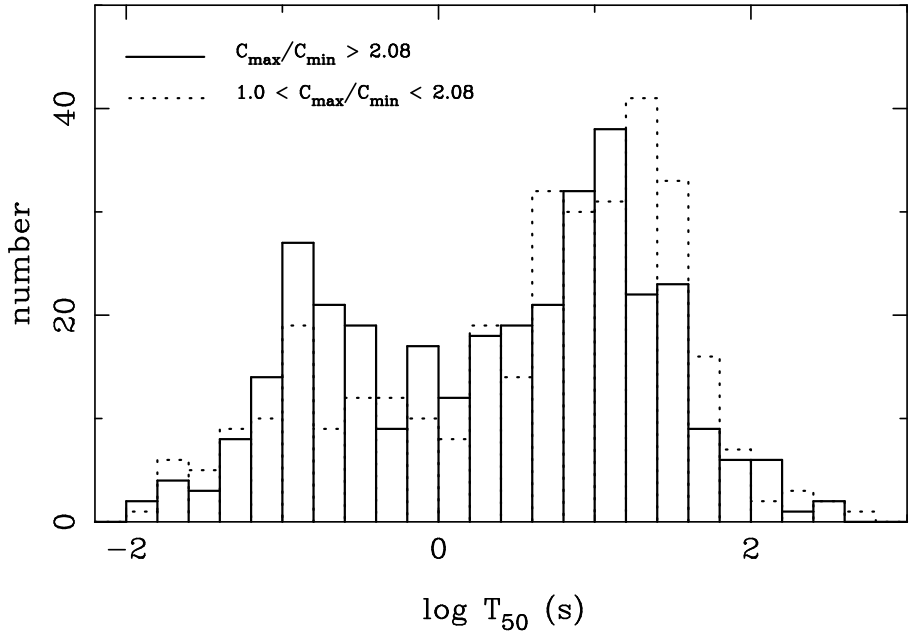}  
\caption{Histogram of durations $T_{50}$ for 665 GRBs in the BATSE
catalog, for strong and weak bursts separately.}
\end{figure}

The early work of Norris~\cite{ref:n96} and our simulations involving
$T_{\rm det}$ in the GUSBAD catalog suggest that there is likely
to be a duration bias for weak GRBs in the sense that their
durations are underestimated.  
We show in Figure 3 a histogram of $T_{50}$ durations in the BATSE
catalog, observed while the on-board trigger was in standard mode,
i.e., requiring a $5.5 \sigma$ excess in at least two detectors in the
energy range $50-300$ keV. Such a histogram was used by 
Kouveliotou et al.~\cite{ref:k93} to show that the $T_{50}$ distribution 
is bimodal. The figure shows separate histograms for bursts with values of 
$C_{\rm max}/C_{\rm min}(64 \rm ms)$ larger and smaller than the median
value 2.08. If the duration bias existed in the
BATSE catalog, the weaker bursts should show a larger fraction of short
bursts than the stronger bursts. We do not see the effect in Figure 3;
in fact there is marginal evidence for the opposite effect.

This agrees qualitatively with the data in Figure 1: the bursts that 
have a $T_{\rm det}$ of $1-2$ s, almost all of which are weak, have a 
wide range of $T_{90}$, up to hundreds of seconds. A remarkable
example is the second burst in the BATSE catalog 4B 910423, a weak
burst with $C_{\rm max}/C_{\rm min}(1024 \rm ms) = 1.11$ with a
listed duration of $T_{90} = 208.6 \pm 1.1$ s. Given the results
from the simulations illustrated in Figure 2 and the S/N ratios
quoted there, it is hard to understand how this can be realistic.

The situation for $T_{50}$ is less clear. It is not possible to
convert the S/N ratios around $3-4$ found in the simulations into
a quantitative estimate of the uncertainty in $T_{50}$. Since
$T_{50}$ is the basis for the duration bimodality of GRBs, it would
be particularly useful if the experience (including the use of 
analytical interpolations of the background) that went into the derivation
of durations for the BATSE catalog could be applied to the results
of simulations such as we have described above. In this fashion one
might hope to gain quantitative information about the statistical and 
sytematic errors in $T_{50}$ and $T_{90}$ as a function of peak flux.

\end{document}